\documentclass[12pt]{iopart}
\usepackage{epsf}
\begin{document}

\title[Strong-coupling expansion of the lattice $\phi^4$ model]
{
Strong-coupling-expansion analysis
of the false-vacuum decay rate
of the lattice $\phi^4$ model in $1+1$ dimensions
}

\author{Yoshihiro Nishiyama}

\address{
Department of Physics, Faculty of Science,
Okayama University,
Okayama 700-8530, Japan}

\ead{nisiyama@psun.phys.okayama-u.ac.jp}

\begin{abstract}

Strong-coupling expansion is performed for the lattice
$\phi^4$ model in $1+1$ dimensions.
Because the strong-coupling limit itself is not solvable,
we employed numerical calculations so as to set up
unperturbed eigensystems.
Restricting the number of Hilbert-space bases, 
we performed linked-cluster
expansion up to eleventh order.
We carried out alternative simulation by means of the
density-matrix renormalization group.
Thereby, we confirmed that our series-expansion data
with a convergence-acceleration
trick are in good agreement with the simulation result.
Through the analytic continuation to the domain of negative biquadratic 
interaction,
we obtain the false-vacuum decay rate.
Contrary to common belief that tunnelling phenomenon lies out of  
perturbative treatments,
our series expansion reproduces the instanton-theory behaviour
for high potential barrier.
For shallow barrier, on the contrary, our result tells that the 
relaxation is no more described by instanton, but the decay 
rate acquires notable enhancement.

\end{abstract}

\pacs{
64.60.My Metastable phases,
03.65.Xp Tunneling, traversal time, quantum Zeno dynamics,
12.38.Cy Summation of perturbation theory,
78.20.Bh Theory, models, and numerical simulation,
}

\submitto{\JPA}

\maketitle

\section{Introduction}

Suppose that a system is placed at a certain metastable state 
surrounded by local potential minimum,
the system would be unstable to decay to a global minimum
assisted by either quantum or thermal fluctuations.
Such processes are called false-vacuum decay and metastability
relaxation, and they
are considered to be non-perturbative in nature.
Hence,
in order to calculate the decay rate (life time),
ingenious treatments have been invented so far 
\cite{Langer67,Coleman85,Fisher67,Gunther80}.
Those treatments rely on semi-classical approximation.
That is, the treatments take into account
quadratic fluctuations around 
the field configuration which extremizes the Euclidean action.
Such field configurations are called
instanton, bounce and (critical) droplet.
Therefore,
those treatments, just like the WKB approximation in wave mechanics,
are not justified for strong fluctuations 
(namely, short life time).
In addition, it is quite
cumbersome to improve the approximation systematically.

In order to compensate the above drawback, first-principle
calculation scheme free from any biased errors would be desirable.
As for discrete variable model (kinetic Ising model),
actually, remarkable {\it tour de force} scheme was invented by 
G\"unther {\it et al.} \cite{Gunther94}.
They introduced the so-called constrained-transfer-matrix method,
which meets nonequilibrium situation.
Then, they carried out extensive numerical calculations of
the transfer matrix.
In consequence, they extracted imaginary part of the free energy, 
which is to be identified as the decay rate.
To the best of our knowledge, it is the first {\it ab initio}
approach to the decay rate in the presence of many-body correlations.
Their result
supports the aforementioned analytic theory based on the
droplet picture.
(Besides this,
Monte-Carlo simulation has been utilized to evolve the relaxation processes
\cite{Stoll77,Rikvold94},
where
the number of Monte-Carlo steps is interpreted as time progression.
Though the interpretation is, in a strict sense, not fully justified,
the simulation result is fairly in accordance with the droplet picture
actually.)

On the contrary, as for continuous-variable model
such as the $\phi^4$ model, the above approach does not apply,
and so far, no attempt at {\it ab initio} calculation 
has been reported.
For quantum-mechanics level ($0+1$ dimension), however,
a number of substantial progresses are made \cite{Drummond82}:
Suzuki and Yasuta obtained a compact expression for the decay (tunnelling) rate
based on the weak-coupling expansion and succeeding Borel resummation
\cite{Suzuki97,Suzuki98}.
They succeeded in calculating the tunnelling rate
beyond instanton calculus.
Alternatively, from the weak coupling expansion, 
Karrlein and Kleinert obtained, remarkably enough,
strong-coupling series by means of the so-called variational perturbation
\cite{Karrlein94}.
Both approaches pursue first-principle calculation scheme 
beyond instanton 
calculus.
As a consequence, these theories clarified how the instanton description
fails for low potential barrier;
true decay rate is suppressed owing to inter-instanton interaction.
At present, 
extension to many-body case appears to be unsuccessful
\cite{Suzuki98}.

The aim of this paper is to investigate the false-vacuum decay rate
for many-body system through series expansion.
We studied 
the lattice $\phi^4$ model in $1+1$ dimensions,
\begin{equation}
\label{Hamiltonian}
{\cal H} = \sum_i 
       \left(
          \frac{1}{2}\pi_i^2 
         +\frac{1}{2} (\phi_i-\phi_{i+1})^2
         +\frac{1}{2}\phi_i^2
         +g \phi_i^4       
       \right)                   .
\end{equation}
with the canonical commutation relations
$[\phi_i,\pi_j]= {\rm i} \delta_{ij}$, $[\phi_i,\phi_j]=0$
and $[\pi_i,\pi_j]=0$.
Note that for $g<0$, the potential is not bounded below, and
renders 
the state $\phi \approx 0$ unstable (false vacuum).
The decay rate due to the quantum fluctuations is our concern.
We will show that in contrast to  
$0+1$ dimension mentioned above, 
the decay rate is {\em enhanced} owing to inter-instanton interaction.

The present paper is organized as follows.
In the next section, we calculate the decay rate
by means of strong-coupling expansion.
We explain methodological details, and check the validity
by means of an alternative simulation. 
In the last section, we summarize the present study.

\section{Results and discussions}

In this section, we will calculate the false-vacuum decay rate
of the model (\ref{Hamiltonian}) through strong-coupling expansion.
To begin with, we will formulate the basis of the expansion.

\subsection{Strong-coupling expansion}

Making use of the rescalings
$\phi \to g^{-1/6} \phi$  
and
$\pi \to g^{1/6} \pi$,
we arrive at the 
expression,
\begin{equation}
{\cal H}=g^{1/3} h ,
\end{equation}
where,
\begin{equation}
\label{strong_coupling_Hamiltonian}
h = \sum_i 
       \left(
          \frac{1}{2}\pi_i^2 
         +\phi_i^4       
+\frac{1}{g^{2/3}}\left(
          \frac{1}{2} (\phi_i-\phi_{i+1})^2
         +\frac{1}{2}\phi_i^2
                 \right)
       \right)             .
\end{equation}
According to the formula, the quadratic potential terms are
regarded as perturbations, and so
 the ground-state energy is expanded in terms of 
the strong-coupling parameter
$\lambda=1/g^{2/3}$;
\begin{equation}
E_{\rm g}=g^{1/3} e_{\rm g} ,
\end{equation}
where,
\begin{equation}
\label{strong_coupling_expansion}
e_{\rm g} = \sum_{n=0} a_n \lambda^n .
\end{equation} 
Note that the unperturbed Hamiltonian $h|_{\lambda=0}$
is biquadratic.
Hence, it is not quite straightforward to perform perturbation 
with respect to this limit.
Here, however, we will manage the perturbation expansion
with the aid of computer calculations.

\subsection{Linked-cluster expansion}

The unperturbed Hamiltonian is a collection of 
independent anharmonic oscillators, and the perturbation introduces
coupling among them.
In such case, the linked-cluster expansion 
is useful to generate
perturbation series.
The linked-cluster expansion
 is a method of, so to speak, computer-aided diagrammatic expansion
\cite{Marland81,Hamer84,Singh88}.
To perform cluster expansion, we should set up 
unperturbed eigensystems nevertheless.
For that purpose, we must diagonalize the Hamiltonian of 
each local anharmonic oscillator
$h_i=\pi_i^2/2+\phi_i^4$.
We carried out the diagonalization in the following way:
(a) An oscillator spans infinite-dimensional Hilbert space.
In order to perform computer simulation,
we need to restrict the number of bases.
For that purpose,
we prepare low-lying $M=400$ states of {\it harmonic} oscillator
with quadratic potential $\Omega^2 \phi^2 /2 $; 
namely, $\{ |n\rangle_\Omega \}$ ($n=0,\cdots,M-1$).
Note that the diagonalization of $h_i$ is now manageable,
because
the Hilbert space is spanned
by finite number of bases just prepared.
Here, $\Omega$ is a freely tunable variational
parameter,
and we had 
adjusted it so as to minimize
${}_\Omega\langle 0 | h_i | 0 \rangle_\Omega$;
 namely, we chose $\Omega=6^{1/3}$.
(This idea is a reminiscence of Feynman and Kleinert \cite{Feynman86}, who
calculated the thermodynamics of anharmonic oscillator
by replacing biquadratic potential with an optimal quadratic one.)
(b) With respect to the Hilbert-space frame 
$\{ |n\rangle_\Omega \}$ ($n=0 \sim M-1$), we represented
the anharmonic-oscillator Hamiltonian $h_i$, and diagonalized
it to obtain the energy levels and the eigenvectors.
(c) Provided that those eigenvectors are at hand,
we carried out secondary Hilbert-space truncation:
We extracted low-lying $m$ eigenvectors among $M$,
and discarded the others.
Henceforth,
those $m$ vectors are to be used to span
the (intra-oscillator) Hilbert space.
(Such Hilbert space restriction scheme originates in Wilson,
who diagonalized
huge cluster of conduction electrons \cite{Wilson75}.)

To summarize, we truncated the intra-oscillator bases through two steps.
First, we had utilized the eigenvectors of a {\em harmonic}
oscillator to span the Hilbert-space frame.
Those are not very efficient, and so, we prepared
rather huge number of $M=400$ bases in practice.
In that sense, the second truncation is significant, where we had
remained only low-lying $m$ bases after solving the eigensystems of 
the intra-site Hamiltonian $h_i$.
These bases turned out to be very efficient (see below), 
and only $m=10\sim25$
bases are necessary so as to achieve reliable calculations 
in the succeeding 
linked-cluster expansion.
(Note that to perform the linked-cluster expansion,
we need to store, in computer memory, huge 
Hilbert-space vector
for clusters consisting {\em many} oscillators.)

Before going into cluster expansion,
we will check the reliability of the Hilbert-space restrictions.
We treat a single anharmonic oscillator
(namely, we ignore the inter-oscillator coupling)
with respect to the restricted Hilbert space mentioned above.
We used the ordinary
Rayleigh-Schr\"odinger perturbation theory,
because the system is of one-body problem.
The strong-coupling perturbation coefficients are 
reported in the literature \cite{Janke95}.
We observed following encouraging features:
First, the choice of $M=400$ is sufficient.
Namely, it reaches the limit of numerical round-off error
(we used extended precision of 16-byte real number), and 
further increase of $M$ just alters final few digits.
Secondly, we found that
rather small $m$ yields precise data.
For example,
$m=15$, which would seem exceedingly small, 
reproduces the perturbation coefficients reported in \cite{Janke95}
with high
precision of order $ \sim 10^{-17}$
(that is not relative but absolute error).
Moreover, the
precision is maintained even for high-order
 perturbation coefficients.
For example, the choice of $M=400$ and $m=25$,
for which
the simulation takes ten minutes or so, is
 sufficient to reproduce
the full result of \cite{Janke95}.

Encouraged by these findings, we performed the linked-cluster expansion 
for the lattice $\phi^4$ model \eref{strong_coupling_Hamiltonian}.
We 
obtained the perturbation series up to eleventh order.
The strong-coupling series is given by,
\begin{eqnarray}
e_{\rm g} &=&
      0.66798625915577710827096201688     
                                                        \nonumber \\
         &  &
     +0.43100635014259473006095738275     
\lambda                                                    \nonumber \\
         &  &
     -0.10148809521111863294125944502     
\lambda^2                                                  \nonumber \\
         &  &
     +0.04803845646443637442034775341     
\lambda^3                                               \nonumber \\
         &  &
     -0.029018513979643624653232757064    
\lambda^4                                                \nonumber \\
         &  &
     +0.019777791330895673863274529570    
\lambda^5                                                \nonumber \\
         &  &
     -0.014454753622894705466341917665    
\lambda^6                                                \nonumber \\
        &  & 
     +0.01106139124598227911409431586     
\lambda^7                                             \nonumber \\
     &  &
     -0.008749346526997    2 
\lambda^8                                            \nonumber \\
     &  &
     +0.00709674759180     5 
\lambda^9                                                  \nonumber \\
     &   &
     -0.00587142           8 
\lambda^{10}
     +0.0049362            2 
\lambda^{11}                     
\label{series_expansion}
\end{eqnarray}
with uncertainties only in the final digits.

\subsection{Resummation and its verification with DMRG}

\begin{figure}
\begin{center}
\epsfbox{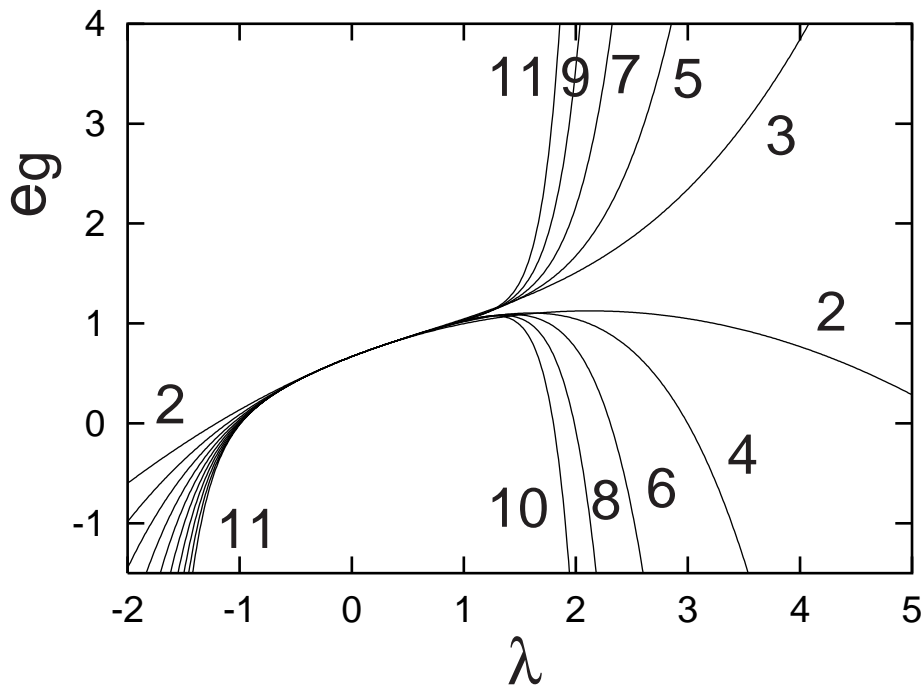}
\end{center}
\caption{
Strong-coupling series $e_{\rm g}(\lambda)$ (\ref{series_expansion}) is plotted.
The series is truncated at various orders.
Sudden deviation may indicate the convergence bound.
Convergence-accelerated data are presented in \fref{figure2}.
}
\label{figure1}
\end{figure}

In the above, we obtained strong-coupling series expansion
for $e_{\rm g}$ (\ref{series_expansion}).
We plotted the result in \fref{figure1}.
We had truncated the series at various orders, which are indicated
for respective curves.
We see that the
curves start to deviate at $\lambda \approx 1$, and
higher-order data exhibit even worse convergence.
Hence,
it is suggested that the series (\ref{series_expansion}) 
has finite convergence radius $|\lambda| \sim 1$.
In order to go beyond the convergence bound and extract meaningful physics,
we have to process our data with some resummation trick.

\begin{figure}
\begin{center}
\epsfbox{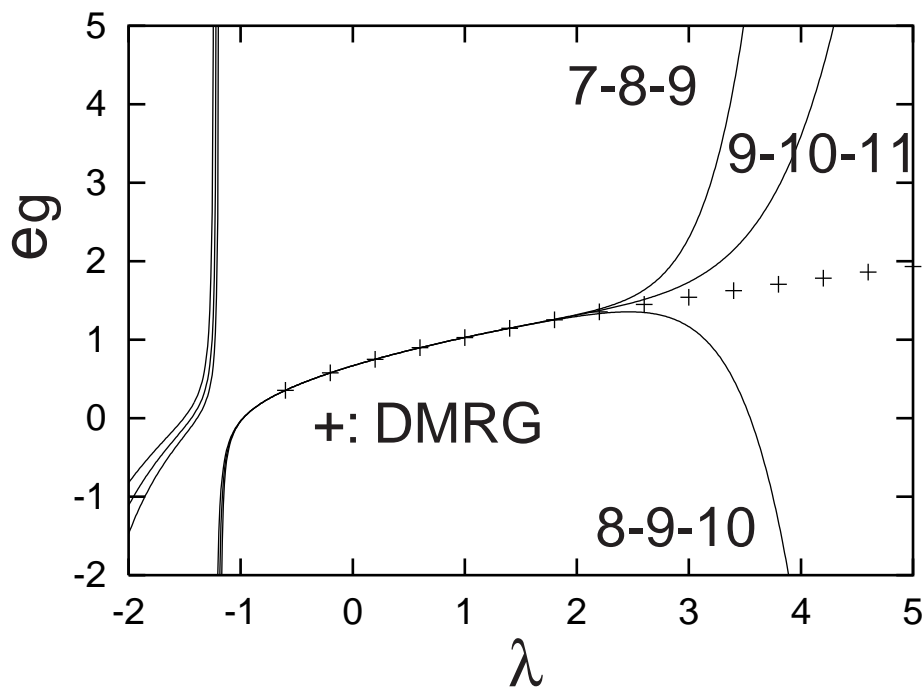}
\end{center}
\caption{The same as \fref{figure1}, but the data are convergence-accelerated
by the formula (\ref{Aitken}).
The symbol such as $l$-$m$-$n$ indicates that the data are 
processed
with use of three partial sums $S_l$, $S_m$ and $S_n$.
We also presented a first-principle simulation result
by means of the density-matrix renormalization group.
We confirm that our series achieves good convergence over 
the range $\lambda<2$.
}
\label{figure2}
\end{figure}

We found that Aitken's $\delta^2$-process \cite{NRF},
\begin{equation}
\label{Aitken}
S'_n=S_{n}-\frac{(S_{n}-S_m)^2}{S_{n}-2S_{m}+S_{l}}                ,
\end{equation}
is very useful to accelerate the convergence of our series.
Here, $S_l$, $S_m$ and $S_n$ are three successive partial sums
truncated at respective orders.
We plotted the resumed results in \fref{figure2}.
The symbol such as $5$-$6$-$7$ indicates that
the data are accelerated with 
the partial sums of $S_5$, $S_6$ and $S_7$.
We see that the data exhibit pronounced convergence improvement.
Our data may be valid up to $\lambda\approx 2$.

In order to check the convergence bound more definitely, 
we performed an
alternative first-principle simulation with the density-matrix
renormalization group.
Our algorithm is standard.
As for a comprehensive overview of this algorithm,
interested readers may consult with  
a proceeding \cite{Peschel99}. 
Full account of technical details specific to the
$1+1$-dimensional scaler field theory will be
found in our paper
\cite{Nishiyama01}.
(In this paper, we studied field $\phi$ confined
within the rigid-wall potential $V(\phi)$.
In order to match the present case, one has to
replace $V(\phi)$
with $\phi^4$.)
The numerical error was checked throughly in \cite{Nishiyama01},
and it was found to reach $10^{-7}$.
We monitored the performance in the present case as well,
and found that
the precision 
is maintained.
The error would be far less than the symbol size 
shown in the plot \ref{figure2}.

The first-principle data are shown in \fref{figure2} as well.
We see that our resummed data are valid
up to $\lambda \approx 2$ fairly definitely.

Finally, we mention a singularity occurring at 
$\lambda\approx -2( <0)$; see \fref{figure2}.
It is noteworthy that for $\lambda<0$, the potential becomes 
double-well form.
Therefore, 
at a certain critical $\lambda$,
there would be an Ising-type phase transition.
The singularity found in our data may indicate the onset of the
transition.
Determination of the critical point for the lattice $\phi^4$ model
is attracting considerable attention recently in the context of 
quantum ferroelectric transition
\cite{Rubtsov01}.
We will pursue this issue elsewhere, and 
in the present paper, we will not go into details any further.

\subsection{Analytic continuation to $g<0$: false-vacuum decay rate}

In the above, we attained good convergence of 
the series expansion (\ref{series_expansion}) with the aid of the 
convergence-acceleration trick (\ref{Aitken}).
Armed by this achievement, in this subsection,
we access the domain of $g<0$
through the analytic continuation $g\to -g$.
For $g<0$, potential is not bounded below, and exhibits a
local potential minimum in the vicinity of $\phi=0$ (metastability).
Because the series expansion (\ref{series_expansion}) is
an irrational function in terms of $g$,
the analytic continuation
renders imaginary part in the ground-state energy.
Thereby, from it, we can read off the false-vacuum decay rate.
In practice, the analytic continuation is done
through the path $g \exp({\rm i}\theta)$ 
of $\theta=0 \to \pi$. 
That is, the term $g^{1/3}$
gives rise to the contribution 
$g/2+{\rm i}\sqrt{3}g/2$ after $g\to -g$.

So far, as to calculate the decay rate,
the instanton technique has been used.
The technique is justified for sufficiently large potential 
barrier (small $g$).
In the following we will show that our series-expansion approach
covers the instanton theory.

In \fref{figure3}, we plotted the false-vacuum decay rate 
multiplied by $g$, namely,
$ g {\rm Im}E_{\rm g}(-g)$,
against $1/g$.
The factor $g$ should kill the prefactor of
a dominant exponential contribution:
That is,
the instanton theory predicts that the
decay rate should obey the formula,
\begin{equation}
\label{instanton}
{\rm Im}E_{\rm g}(-g) \propto \frac{1}{g}\exp(-S/g)   ,
\end{equation}
where
$S$ denotes the Euclidean action of one instanton.
We solved the instanton solution numerically, and obtained the
estimate,
\begin{equation}
S=1.1891027(5)                .
\end{equation}
To be concrete, we will sketch the calculation method.
First of all, one must reformulate the Hamiltonian
formalism (\ref{Hamiltonian}) 
into the Lagrangian formalism in the Euclidean space-time. 
Thereby,
we considered the system with 28 sites and imaginary time $\beta=28$.
The imaginary time is discretized into 8000 intermediate time slices.
(Note that now, the field is defined in the discretized space-time.)
Because the instanton solution (field configuration)
should minimize the Euclidean action,
the problem reduces to the minimization of multi-dimensional
function.
That computation is readily achieved by the utilities
supplied in simulation guide books such as \cite{NRF}.
The amount of error is estimated with changing the system sizes
and discretization intervals.

We would like to draw reader's attention to the fact
that
the formula (\ref{instanton}) has an essential singularity at $g=0$.
That is why we had selected the strategy of approaching from $g\to\infty$
rather than from $g=0$.

Let us turn to the discussion of our result of \fref{figure3}.
As is mentioned above,
the instanton result (\ref{instanton}) is validated for large $1/g$.
As a matter of fact, for $1/g>1$, our data approaches the
instanton prediction; we had drawn the slope of \eref{instanton}.
In this respect, the convergence-acceleration trick (\ref{Aitken}) is crucial
in our study, because it enables us to attain good convergence
up to $1/g \sim 3$, which appears to reach the instanton region.

For $1/g < 1$, on the other hand, our data indicate rapid enhancement
of the decay rate; namely, the curve starts 
to deviate from the instanton prediction.
It is to be stressed that our treatment is 
justified for strong-coupling
limit ($1/g \ll 1$).
Therefore, it is found that the inter-instanton correlation
gives rise to {\em enhancement} of relaxation.
This feature is to be contrasted with that of
 $0+1$ dimension, where the inter-instanton 
correlation results in {\em suppression} of decay rate.
Enhancement in $1+1$ dimensions was speculated in the former study \cite{Suzuki98},
where the authors utilized the {\em weak}-coupling expansion and
the Borel technique.
Although their series does not show any indication of convergence,
their result actually captures a signal of relaxation enhancement.

According to Kleinert, in the regime $g \gg 1$, the decay process is 
governed by `sliding' rather than instanton \cite{Kleinert93}.
Nevertheless, we stress that the present series-expansion approach
covers both instanton ($1/g>1$) and sliding ($1/g<1$)
regimes in a unified way.
Moreover,
Our series is readily improved systematically just
by performing cluster expansion further.

In the above, we found that at $1/g \approx 1$, there exists a
crossover boundary separating two distinctive regimes.
Our result supports the previous proposal of \cite{Suzuki98}.
The authors calculated the effective potential,
and found that for
$g > 1.17$, the potential barrier is smeared out by quantum fluctuations.
Their criterion would be sensible for
separating instanton and sliding phases.

In \fref{figure3},
we see that the data of 9-10-11 and 8-9-10 show poor convergence.
That may possibly be due to the fact that our 10th and 11th
perturbation coefficients have rather few significant figures available.

Finally, we recollect past findings for the $\phi^4$ theory
in {\it continuous} space time.
Br\'ezin and Parisi completed the instanton calculation, and obtained the
formula
${\rm Im}E_{\rm g}(-g)=(0.0815435/g)\exp(-1.4626121/g)$ \cite{Brezin78}.
We notice that the instanton action 
is similar to that of our lattice model.
Perhaps, the decay process would be identical between the lattice model
and the
continuous-field theory.

\section{Summary and discussions}

So far, several {\it ab initio} approaches have been proposed
in order to calculate the decay rate beyond semiclassical approximation.
In particular, the $\phi^4$ model in $0+1$ dimension has come
under through investigation \cite{Drummond82,Suzuki97,Suzuki98,Karrlein94},
 while the extension to many-body case
remains unsuccessful.
In the present paper, by means of the strong-coupling expansion,
we studied 
the $(1+1)$-dimensional lattice $\phi^4$ model (\ref{Hamiltonian}).
We demonstrated that the linked-cluster expansion method works
very efficiently,
provided that Hilbert-space restriction is processed
properly.
In addition, we found that the convergence-acceleration trick 
(\ref{Aitken}) 
is significant.
In fact, the convergence-accelerated sum reproduces the first-principle data
for considerably wide range $\lambda < 2$.

\begin{figure}
\begin{center}
\epsfbox{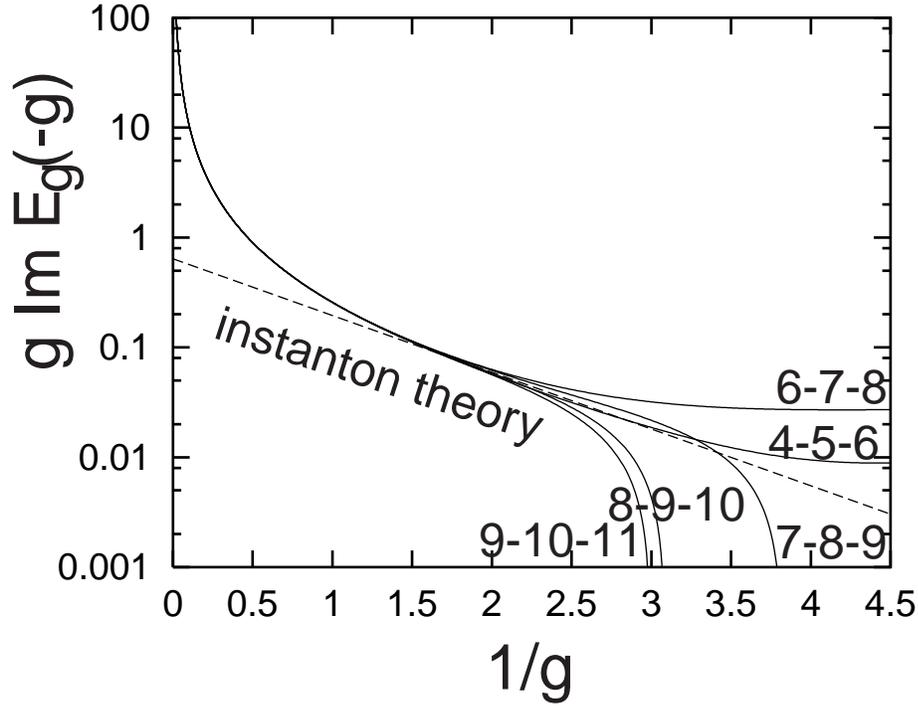}
\end{center}
\caption{
False-vacuum decay rate (multiplied by $g$)
 $g{\rm Im}E_{\rm g}(-g)$ is plotted.
The symbol $l$-$m$-$n$ indicates that the 
data are convergence-accelerated
with use of three partial sums $S_l$, $S_m$ and $S_n$.
We plotted a slope $\exp(-S/g)$ which is predicted
by the instanton theory; see text.
Note that
the instanton treatment is justified for large $1/g$.
As a matter of fact,
our series expansion obeys the prediction for $1/g>1$.
For $1/g<1$, on the contrary, 
our result exhibits notable enhancement.
Hence, it is suggested that the inter-instanton interaction rather
{\em enhances} the relaxation.
}
\label{figure3}
\end{figure}

Based on the above achievements,
we surveyed the domain of metastability
through the analytic continuation $g \to -g$.
We are concerned in the false-vacuum decay rate ${\rm Im}E_{\rm g}(-g)$;
see \fref{figure3}.
Our result indicates that there are two regimes.
For $g<1$, our result obeys the prediction by the instanton theory.
It is to be stressed that the convergence acceleration (\ref{Aitken}) is
significant to reach the instanton regime.
For shallow potential barrier $g>1$,
 the relaxation is no more described by instanton,
but the relaxation rate acquires notable enhancement.
According to Kleinert, for $g>1$, the relaxation is driven by
sliding rather than instanton.
Nevertheless,
we stress that our series-expansion approach does cover both
regions with a unified framework,
and it is readily improved systematically just by continuing the
perturbation further. 
It would be promising that the present method is applied to
other wide class of metastable systems.

As is mentioned above, we had performed 
the density-matrix-renormalization-group simulation as well.
From the simulation data, we are able to
 extract perturbation coefficients
by polynomial fitting.
This technique is applicable to those models
that even possess complicated interactions and spatial inhomogeneity.
Tunnelling phenomena assisted by an impurity is of current interest
\cite{Kato00}.
However, ambiguities in estimating fitting errors
are not fully resolved at present.
It is left for future study.

\end{document}